\documentclass[epj]{svjour}
\usepackage{latexsym}
\usepackage{graphics}
\usepackage{bm}
\newcommand{\ket}[1]{|#1\rangle}

\begin{document}
\title{Trapping and coherent manipulation of a Rydberg atom on a
microfabricated device: a proposal}
\author{John Mozley\inst{1} \and Philippe Hyafil\inst{1} \and Gilles
Nogues\inst{1} \and Michel Brune\inst{1} \and Jean-Michel Raimond\inst{1} \and
Serge Haroche\inst{1}\inst{2}% etc
% \thanks is optional - remove next line if not needed
%\thanks{\emph{Present address:} Insert the address here if needed}%
}                     % Do not remove
\offprints{G. N., \texttt{gilles.nogues@lkb.ens.fr}}

\institute{Laboratoire Kastler Brossel, D\'epartement de physique de l'ENS, 24
 rue Lhomond F-75231 Paris cedex 05\and Coll\`ege de France, 11 place Marcelin
Berthelot F-75231 Paris cedex 05}
\date{Received: date / Revised version: date}
% The correct dates will be entered by Springer
%
\abstract{ We propose to apply atom-chip techniques to the trapping
of a single atom in a circular Rydberg state. The small size of
microfabricated structures will allow for trap geometries with
microwave cut-off frequencies high enough to inhibit the spontaneous
emission of the Rydberg atom, paving the way to complete control of
both external and internal degrees of freedom over very long times.
Trapping is achieved using carefully designed electric fields,
created by a simple pattern of electrodes. We show that it is
possible to excite, and then trap, one and only one Rydberg atom
from a cloud of ground state atoms confined on a magnetic atom chip,
itself integrated with the Rydberg trap. Distinct internal states of
the atom are simultaneously trapped, providing us with a two-level
system extremely attractive for atom-surface and atom-atom
interaction studies. We describe a method for reducing by three
orders of magnitude dephasing due to Stark shifts, induced by the
trapping field, of the internal transition frequency. This allows
for, in combination with spin-echo techniques, maintenance of an
internal coherence over times in the second range. This method
operates via a controlled light shift rendering the two internal
states' Stark shifts almost identical. We thoroughly identify and
account for sources of imperfection in order to verify at each step
the realism of our proposal.
 % end of PACS codes
} %end of abstract
\maketitle
\section{Introduction}\label{sec:intro}

In recent years we have witnessed convergence between many fields of
physics previously considered disparate. This has been particularly
true in the case of solid state and atomic phy\-sics. On one side,
the possibility of designing nanometric devices and the
understanding of quantum phenomena in devices such as Josephson
junctions has allowed experimentalists to control the quantum
coherence of semi-~\cite{QI_SISPINYAMAMOTO05} or
super-conducting~\cite{QI_QUANTRONIUM02,QI_NAKAMURABOX02} systems,
exactly as was done many years ago for atoms or molecules. On the
other, atomic physicists have been able to produce Bose-Einstein
condensates~\cite{MX_CORNELLBEC95} and degenerate Fermi
gases~\cite{MX_FERMIJIN99} in quasi ideal situations, wonderful
tools for probing many body theory and mesoscopic physics.

Atom-chip experiments~\cite{SCHMIEDMAYERREVIEW_99,MX_HANSCHCHIP99}
open the way to investigations at the frontier of these two fields.
Aside from their relative ease of use and potential as atom
interferometers, they provide an ideal environment in which atomic
ensembles can be integrated with, and coupled to, devices on a
microcircuit. Moreover, they allow for the study of atom-surface
interactions with a completely new degree of control. Separately,
experiments on circular Rydberg atoms have proven them extremely
sensitive probes of electric and magnetic fields, both static and
dynamic, making ideal tools for the investigation of numerous
quantum phenomena, atom-surface interactions included. This
remarkable sensitivity, in conjunction with their relative ease of
detection, makes a single circular Rydberg atom an excellent probe
of microwave field intensities ranging from one to a few tens of
photons~\cite{ENS_RMP}. The coupling of a single such atom to a high
Q superconducting microwave cavity has allowed our group to study,
for example, the decoherence of mesoscopic ensembles of
photons~\cite{ENS_CAT} and to realize elementary quantum logic
operations~\cite{ENS_GHZ}. In these experiments, circular Rydberg
atoms are excited from a thermal atomic beam (mean velocity 350~m/s)
crossing a cryogenic setup in which the superconducting cavity is
mounted. The use of thermal atoms intrinsically limits the maximum
interaction time between the atom and the microwave field. Moreover,
in order to reduce to a negligible level the proportion of unwanted
events in which we have more than one circular Rydberg atom present
at once in the cavity, we are forced to work with a very low rate of
excitation towards the Rydberg levels. Aside from increasing
dramatically the acquisition time, this lack of a deterministic atom
source is an inconvenience from the point of view of quantum
information processing.

The integrated trapping and manipulation of circular Rydberg atoms
on a chip proposed here would lift these limitations. If we are to
take advantage, however, of the longer interaction times achievable
with such trapped Rydberg atoms, we must also combine in the same
setup the capability for the extension of the natural lifetime and
long term coherent manipulation. For short term manipulation over
times of the order of the millisecond, this being much shorter than
their lifetime in free space, one could simply excite the Rydberg
atoms from an initial sample of cold trapped atoms and study them in
free fall~\cite{QI_GRANGIERRYD02}. In Ref.~\cite{ENS_TRAPRYDBERG04}
we proposed an electric trap able not only to store an individual
circular Rydberg atom, but also to inhibit its spontaneous emission
and maintain an internal coherence. In this paper we present in much
more detail how we assess the trap's performance, how to integrate
it on a chip and how we simulate the different imperfections of the
setup. We not only present the trap geometry of
Ref.~\cite{ENS_TRAPRYDBERG04} but also show that it is possible to
extend the proposal to a smaller design where the atom is brought
closer to the chip surface, a design therefore better adapted to
atom-surface interaction studies. The principle of the trap, along
with its implementations in realistic geometries, is discussed in
Sec.~\ref{sec:trap}. We show in Sec.~\ref{sec:loading} that the
excitation of a sample of ground state atoms, initially trapped on a
magnetic atom chip, can lead to the preparation of a single Rydberg
atom, taking advantage of the dipole-blockade
effect~\cite{QI_LUKINDIPOLEBLOCKADE01}. As explained above, the
maximum duration of an experiment with this atom is not limited by
the storage time but by the lifetime of the circular Rydberg state.
We discuss in Sec.~\ref{sec:spontemission} the efficiency of the
spontaneous emission inhibition when we place our atom chip within a
structure excluding the millimetre-wave decay channel. We show that
the radiative lifetime could realistically be pushed into the second
range. We present finally in Sec.~\ref{sec:dephasing} a technique
taking advantage of a controlled light shift of the Rydberg levels
allowing for the maintenance and control over a similar duration of
the coherence of an atom in a superposition of two trapped states.

\section{A microfabricated electrodynamic trap for Rydberg
atoms}\label{sec:trap}

The large coupling of Rydberg atoms to static or time-varying fields
presents us with numerous possible trapping techniques. It has been
proposed, for example, to  use powerful, far-detuned laser beams to
create very tight traps for Rydberg states. These traps exploit the
ponderomotive force~\cite{TR_PONDEROMOTIVE00} experienced by the
atom due to the fast oscillation of its valence electron in the
laser field. It remains hard however to fulfill all the criteria
allowing for coherent manipulation of Rydberg levels in any such
trap. For instance, one can see that the very large laser power
required by the ponderomotive trapping is hardly compatible with the
cryogenic environment necessary to avoid blackbody radiation in the
microwave domain~\cite{ENS_RMP} and the consequent destabilization
of the Rydberg levels.

Another possible solution is to use d.c. magnetic fields and
conventional atom-chip architectures. The magnetic moment of a
Rydberg level (principal quantum number $n$) can be of the order of
$n \mu_B$, where $\mu_B$ is the Bohr magneton. The same atom-chip
wires could be used for the trapping of ground and Rydberg state
atoms. One possible problem is that the Rydberg state might be
affected by its interaction with still-trapped ground-state atoms in
its vicinity. Moreover, the transition frequency between different
Rydberg states is strongly perturbed by the Zeeman effect. We could
not find a way to suppress the resulting random dephasing, in
contrast to that due to the Stark effect, which can be partially
compensated as explained in Sec.~\ref{sec:dephasing}.

In the following we therefore use time-varying electric fields to
trap the Rydberg atoms.

\subsection{Circular Rydberg atoms in an electric field}\label{ssec:cra}

We consider the case of Rydberg atoms in circular states. These have
a large principal quantum number $n$, of the order of 50 in our
discussion, along with maximal angular and magnetic quantum numbers
($l=|m|=n-1$). The atom therefore has both a large energy and a
large angular momentum. The classical analog of the electron's
wavefunction is a Keplerian circular orbit around the nucleus. To be
concrete we will consider ${}^{87}$Rb atoms. To a very good
approximation one can neglect the fine structure corrections to the
circular levels' energies and assume that they are equal to those of
Hydrogen. We will label $\ket{g}$ the circular Rydberg state with
principal quantum number $n=50$ and $\ket{e}$ that with $n=51$. In
the presence of an electric field the good quantum numbers are $n$,
$m$ and the parabolic number $n_1$~\cite{TXT_GALLAGHER}. The
circular state corresponds to the case $|m|=n-1$, $n_1=0$ (it is
simultaneously an eigenfunction in the $\{n,l,m\}$ and $\{n,n_1,m\}$
representations) and the normal to the plane of its circular orbit
aligns parallel to the electric field. Due to the Stark effect the
degeneracy of the manifold of levels of equal $n$ but differing
$n_1$ and $m$ is lifted. Under these conditions the radiative
lifetime of the levels $\ket{g}$ and $\ket{e}$ is extremely long,
$\tau_{sp}\approx30$~ms. For a given electric field amplitude $E$
the energy of each level can be calculated using either a direct
diagonalization of the Stark Hamiltonian (see
Sec.~\ref{ssec:dressingimperf}) or a perturbative approach. Using
the latter method, to second order, the energy $\mathcal{E}(E)$ of a
given level $\ket{n,n_1,m}$ is given by:
\begin{eqnarray}
    \mathcal{E}(E)    & = & \mathcal{E}^{(0)} + \mathcal{E}^{(1)} +
                            \mathcal{E}^{(2)}, \label{eq:stark0-2} \\
    \mathcal{E}^{(0)} & = & -\frac{1}{2n^2},  \\
    \mathcal{E}^{(1)} & = & -\frac{3}{2}(n-2n_1-|m|-1)nE,\\
    \mathcal{E}^{(2)} & = & -\frac{1}{16} [17n^2-9m^2+19 \nonumber \\
                      &   & -3(n-2n_1-|m|-1)^2 ]n^4E^2,
\end{eqnarray}
where all quantities are expressed in atomic units (Energy:
$\mathcal{E}_{a.u.}=4.360 \times 10^{-18}$~J, electric field:
$E_{a.u.}=514.2 \times 10^{9}$~V/m). In the presence of a
time-varying electric field, the normal to the plane of the orbit
follows the field direction so long as the characteristic frequency
of the change in electric field remains very small compared to the
transition frequency $\nu_{\mbox{\tiny{Stark}}}=\mathcal{E}^{(1)}/h$
between the neighbouring non-circular levels. In an electric field
of 400~V/m, typical of the fields considered in this article, this
frequency is of the order of 400~MHz, much higher than all the
trapping frequencies found later, and therefore than the rate of
change in angle of the field. We can therefore in all the following
consider the evolution in the time-varying field to be adiabatic and
that the atom remains in the circular Rydberg state~\cite{ENS_DAHU}.

\begin{figure}
\begin{center}
\resizebox{0.85\columnwidth}{!}{
\includegraphics{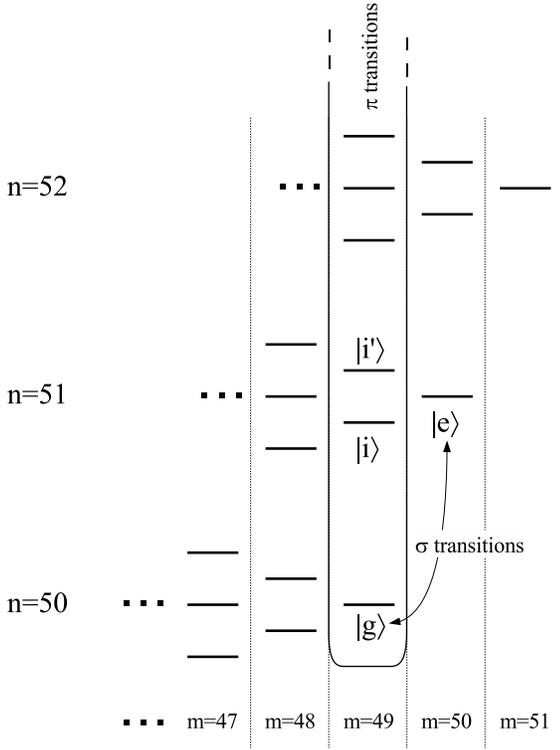}
}
\end{center}
\caption{\label{fig:Stark} Level spectrum in presence of an electric
field for the multiplicities $n=50,51$ and $52$. Only the levels
with the largest angular momentum are presented ($m\geq n-3$). A
$\pi$-polarized microwave field will couple levels of equal $m$, for
example those of $m=49$ separated from the others by a box in this
figure. A $\sigma$-polarized microwave field will, in contrast,
couple levels of $m$ differing by $\pm 1$, for example levels
$\ket{g}$ and $\ket{e}$.}
\end{figure}

Fig.~\ref{fig:Stark} presents the energy levels in the presence of
an electric field for the principal quantum numbers $n$ relevant to
our discussion. It is important to note that circular states, having
no permanent mean electric dipole, experience no linear Stark effect
($\mathcal{E}^{(1)}=0$). They are, however, much more highly
polarizable than ground state atoms and have an accordingly large
quadratic Stark effect [$\mathcal{E}_i \approx \alpha_i E^2$ for
$i=(g,e)$, where $\alpha_i$ is half the polarizability]. Its value,
$\alpha_g=-203,2$ Hz/(V/m)$^2$ for $n=50$ and
$\alpha_e=-228,7$~Hz/(V/m)$^2$ for $n=51$, is 9 orders of magnitude
larger than that of the ground state of Hydrogen. Nonetheless,
circular states are high-field seekers and cannot therefore be
trapped by any configuration of d.c. electric fields, a maximum of
the electric field modulus in vacuum being forbidden by Max\-well's
equations.

\subsection{Trapping}\label{ssec:trapping}

A similar situation is encountered in the case of charged particles,
such as ions. In that particular example, Max\-well's equations
prevent us from obtaining an extremum of the electric potential in a
region devoid of charge. It is nevertheless possible to trap ions if
one uses a.c. electric potentials, as in the case of Paul
traps~\cite{PAULNOBEL_80}. Building on these ideas it has been shown
by Peik~\cite{TR_PEIK99} that polarizable atoms can be trapped if
one combines a relatively strong, d.c. and homogeneous electric
field $\vec{E}_1=E_1 \vec{u}_z$ (where $\vec{u}_z$ is a unit vector
along the vertical) and an inhomogeneous a.c. field
$\vec{E}_3(\vec{r},t)$ deriving from a time-varying hexapolar
potential. As a first step we consider the case where $E_3 \ll E_1$.
The field is therefore almost homogeneous over the trapping region
and small departures from $\vec{E}_1$ are responsible for the
confinement. Let $V_1(\vec{r})$ and $V_3(\vec{r},t)$ be the
potentials associated to $\vec{E}_1$ and $\vec{E}_3(\vec{r},t)$
respectively. One can write these potentials in the form:
\begin{eqnarray}
V_1(x,y,z) & = & \frac{U_1 z}{z_0}, \\
V_3(x,y,z,t) & = & \frac{U_3(t)}{2z_0^3}(2 z^3-3zx^2-3zy^2),
\end{eqnarray}
where the $U_i$ are quantities homogeneous to electric potentials
and $z_0$ is a length of the order of the electrode size. If one
considers a circular Rydberg atom following adiabatically the
electric field variations, its potential energy in the quadratic
Stark approximation is given by:
\begin{eqnarray}
\mathcal{E} & = & \alpha |\vec{E}_1+\vec{E}_3|^2 \nonumber \\
& = & \alpha ( E_1^2 + 2 \vec{E}_1 \cdot \vec{E}_3 + E_3^2).
\end{eqnarray}
Assuming $E_3\ll E_1$ and neglecting the constant terms in the
potential energy we arrive at:
\begin{eqnarray}
  \mathcal{E} & \simeq & 2 \alpha E_1 \vec{E}_3 \cdot \vec{u}_z
\nonumber \\
    & \simeq &  2 \alpha \frac{U_1}{z_0} \frac{\partial V_3(x,y,z,t)}{\partial
    z}\nonumber \\
    & \simeq &  \alpha \frac{3U_1U_{3}(t)}{z_0^4} (2z^2-x^2-y^2).\label{eq:idealpotquad}
\end{eqnarray}
We see that we now have an expression for the potential energy
quadratic in $x$, $y$ and $z$, as in conventional ion traps. It is
important to note that if we have trapping along directions $x$ and
$y$ (i.e. $U_1 U_3(t) > 0 \mbox{ for } \alpha < 0$) then we
necessarily have antitrapping along direction $z$, and vice versa.
Moreover, if $U_3$ varies sinusoidally [i.e. $U_3(t)=U_{30} \cos
(\omega t)$] then the equation of motion along $Ox$ can be written:
\begin{equation}\label{eq:motion_x}
  m_{Rb} \frac{d^2 x}{d t^2}-  \frac{6 \alpha U_1
U_{30}}{z_0^4}\cos(\omega t) x = 0.
\end{equation}
Equation \ref{eq:motion_x} can be transformed into the well known
Mathieu form:

\begin{equation}\label{eq:mathieu}
  \frac{d^2 x}{d\tau ^2}-2 q_x \cos (2\tau)x =0,
\end{equation}
where $\tau=\omega t /2$ et $q_x=12 \alpha U_1 U_{30}/m_{Rb}
\omega^2 z_0^4$.

The equation of motion is of the same form along the other
directions with $q_y=q_x$ and $q_z=-2q_x$. There exists a stable
solution, confined in space, if $|q_i| \leq 0.907,\ i=x,y,z$.

In order to compensate the force of gravity, taken to be along $-z$,
it is also possible to add to $V_1$ and $V_3$ a d.c. quadrupolar
potential $V_2(\vec{r})=U_2(x^2+y^2-2z^2)/z_0^2$. If $U_2 \ll U_1$,
the same analysis as before, considering only the terms in $U_1$ and
$U_2$, leads to:
\begin{eqnarray}
\mathcal{E} & \simeq &  2 \alpha \frac{U_1}{z_0} \frac{\partial
V_2(x,y,z)}{\partial z} \nonumber \\
 & \simeq & -4 \alpha \frac{U_1 U_2}{z_0^3} z,
 \label{eq:Epotidealgrav}
\end{eqnarray}
compensating gravity if $U_2 = mgz_0^3/4\alpha U_1$.

\subsection{Proposed geometries, calculation of
potentials and fields}\label{ssec:trapgeometry}

\begin{figure}
\begin{center}
\resizebox{0.85\columnwidth}{!}{
\includegraphics{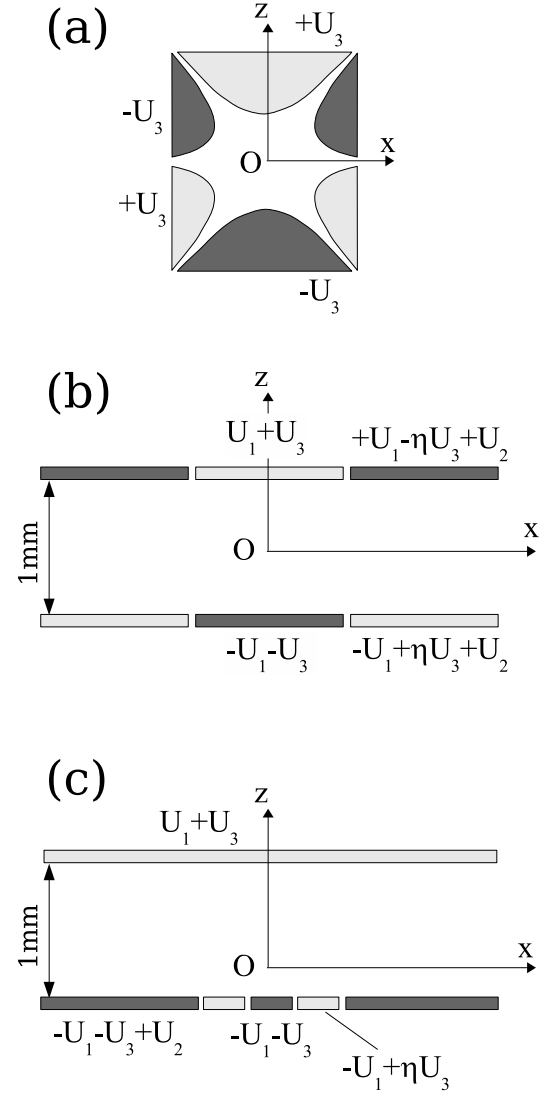}
}
\end{center}
\caption{\label{fig:TrapGeometry} For all figures there is a
cylindrical symmetry around $Oz$ and the electrodes are shaded
according to the phase of the oscillating potential $U_3$. (a)
Electrode geometry creating an exact hexapolar potential. (b)
Section of trap A in a vertical $xOz$ plane, with applied
potentials. The diameter of the inner electrode is 1~mm. The plate
spacing, 1~mm, is appropriate for spontaneous emission inhibition.
The analogy with the geometry of (a) is conspicuous. (c) Geometry of
trap B. The lower plane is not drawn to scale, the inner disk
electrode having a diameter of 100~$\mu$m and the ring around it
being of width 100~$\mu$m. The plate spacing remains 1~mm.}
\end{figure}

It is of course possible to design a geometry of electrodes which
perfectly maps the desired hexapolar geometry [see Fig.
\ref{fig:TrapGeometry} (a)]. It requires two rings and two end caps.
However, being closed in all three dimensions,  such a geometry is
hardly compatible with efficient atom loading, nor with the idea of
building all the trapping components on a microfabricated structure.
We propose instead a design (trap A) based on two chips facing each
other [see Fig. \ref{fig:TrapGeometry} (b)]. Each chip could be made
by standard lithographic methods and consists of two electrodes: a
disk, playing the role of end cap, surrounded by a plane. The
diameter of the disks is 1~mm and the outer electrodes would extend
to $y=\pm 1\mbox{~cm}$ and even further in $x$, the electrodes
forming part of a long waveguide (see
Sec.~\ref{ssec:leveldressing}). To ease the numerical calculation of
the potentials we cut the electrodes off at $\rho=\sqrt{x^2+ y^2}=
2\mbox{~mm}$ and confirmed afterwards that this had a negligible
effect upon the potential within the trapping region.

The voltage applied to each electrode is the sum of 3 terms:
\begin{itemize}
\item A static voltage $\pm U_1$ creating a homogeneous field
$\vec{E}_1$ along $Oz$, as in a plane capacitor.
\item A static voltage $U_2$ on the outer ring of both chips
creating a potential with a large quadrupolar component of the same
form as $V_2(\vec{r})$. This component, in conjunction with
$\vec{E}_1$ as seen above, will create a force on the atom
compensating gravity.
\item A time-varying voltage $U_3(t)$ creating a potential with
a large hexapolar component of the same form as $V_3(\vec{r},t)$.
This component, in conjunction with $\vec{E}_1$ as seen above,
allows for trapping.
\end{itemize}

It is, however, important to note that the electric potential
associated to $U_2$ (resp. $U_3$) is clearly not an ideal quadrupole
(resp. hexapole). The resulting electric fields are therefore more
complicated. As an illustration, if the factor $\eta$ of Fig.
\ref{fig:TrapGeometry} (b) is set equal to 1, $U_3$ will create a
non-zero electric field along $Oz$ at the centre of the trap $O$.
This will create a time-varying electric field, and therefore atomic
energy, at trap centre, resulting in heating. A correct choice of
$\eta$ allows us to better approximate a perfectly hexapolar
potential, and cancel this electric field.

More specifically, our final goal is to be able to calculate the
electric field at any position inside the trap and at any time.
Moreover, it is important to assess edge effects as well as the
effects of the finite gap between the electrodes. We have therefore
opted for a numerical approach. We numerically calculate the
potential inside the trap on a grid of points using the software
SIMION \cite{SIMION}. In order to limit numerical errors in the
calculation of the gradient we then make a fit of the potential on
the basis of the first seven spherical harmonics
$Y_l^m(\theta,\phi)$, centred on $O$. Due to the cylindrical
symmetry around $Oz$ only the $m=0$ terms are present. Thanks to the
linearity of Maxwell's equations one can add separately the
contribution of each voltage $U_k\ (k=1,2,3)$ and we therefore have,
neglecting the physically unimportant constant term:
\begin{equation}
V_{\mbox{\tiny{Fit}}}(r,\theta,\phi,t)=\sum_{l=1}^{7} \sum_{k=1}^{3}
v_{kl} U_k(t) \left ( \frac{r}{z_0} \right )^l Y_l^0(\theta),
\end{equation}
where $z_0$ is an arbitrary length, chosen to be characteristic of
the trap geometry, and set equal to 1~mm for trap A. The parameters
$v_{kl}$ only depend on the geometry of the electrodes and are
determined from our fit on the numerically calculated grid. Due to
symmetries one can set $v_{1l}=0$ for $l\geq 2$, $v_{2l}=0$ for $l$
odd and $v_{3l}=0$ for $l$ even. The results of the fits are
presented in Table~\ref{tab:fitharmonique}. The relative difference
between the results of SIMION and the fit is smaller than 1\% within
a radius of 400~$\mu$m from $O$.
\begin{table*}
\begin{center}
\begin{tabular}{c|ccccccc}
\hline\noalign{\smallskip}
$k \backslash l$ & 1 & 2 & 3 & 4 & 5 & 6 & 7  \\
\noalign{\smallskip}\hline\noalign{\smallskip}
$1$ &  4.09     & 0       & 0       & 0        &  0      &  0      & 0  \\
$2$ &  0        & -3.12   & 0       & -0.63    &  0      &  3.73   & 0  \\
$3\ (\eta=1)$ &  2.60        & 0       &  5.48   & 0        &  -3.66
 &  0      & -5.75 \\
$3\ (\eta=4.49)$ & 0     & 0       & 15.04       & 0        &  -10.05      &
0      & -15.76 \\
\noalign{\smallskip}\hline
\end{tabular}
\end{center}
\caption{\label{tab:fitharmonique}       % Give a unique label
Results of the fit of the potential coefficients $v_{kl}$ for trap A
($z_{0}=1 \mbox{~mm}$). The first two lines give values for the fit
of applied voltages $U_1$ and $U_2$ respectively, the third line for
applied voltage $U_3$ with $\eta=1$. It is possible to cancel the
electric field created by $U_3$ at the centre of the trap by setting
$\eta=4.49$ (fourth line)}
\end{table*}

We have also designed another trap geometry (trap B) which achieves
tighter confinement and smaller atom-electrode distances [see
Fig.~\ref{fig:TrapGeometry}(c)]. Such a setup is necessary for
atom-surface interaction studies. Trap B is derived from trap A by
bringing the ring of the top plane of Fig.~\ref{fig:TrapGeometry}(b)
down onto the lower plane. The electrode size is reduced by a factor
of 10 (and $z_0$ accordingly reduced to $100~\mu$m) while the
distance between the two planes remains unchanged at 1~mm. Numerical
calculations with SIMION show that there exists a saddle point for
the electric field modulus at a point $O$, 120~$\mu$m above the
lower plane surface. It is possible to apply the same numerical
treatment to this trap as before and fit the electric potential
around $O$ (here however we have fewer symmetries and can
consequently set fewer of the $v_{kl}$ equal to zero). This
procedure produces a fit correct to 0.1\% within a radius of
50~$\mu$m of $O$. The results are shown in
Table~\ref{tab:fitharmoniqueMicroTrap}.

\begin{table*}\sidecaption
\begin{tabular}{c|ccccccc}
\hline\noalign{\smallskip}
$k \backslash l$ & 1 & 2 & 3 & 4 & 5 & 6 & 7  \\
\noalign{\smallskip}\hline\noalign{\smallskip}
$1$ & 0.409     & 0       & 0       & 0        &  0      &  0      & 0\\
$2$ & 0.457    & -0.220   & 0.037  & 0.010   &  -0.010 & 0.001 & 2$\times$10$^{-6}$   \\
$3\ (\eta=1)$ & -0.369 & 0.002   &  0.250 & -0.241    & 0.155  &
-0.075
& -0.027 \\
$3\ (\eta=0.05)$ & 0 & 0.001 & 0.131 & -0.127 & 0.082 & -0.039 & -0.014 \\
\noalign{\smallskip}\hline
\end{tabular}
\caption{\label{tab:fitharmoniqueMicroTrap}       % Give a unique label
Results of the fit of the potential coefficients $v_{kl}$ for trap
B ($z_{0} = 100 ~\mu \mbox{m}$). The different lines have the same
meaning as in Table~\ref{tab:fitharmonique}.}
\end{table*}

\subsection{Simulation of trajectories}\label{ssec:trapsimulations}

The evolution of the atom in the time-varying electric field being
adiabatic (see Sec.~\ref{ssec:cra}) we can simply write
$\mathcal{E}(t)=\mathcal{E}[E(t)]$. From $V_{\mbox{\tiny{Fit}}}$, it
is easy to derive a formula for the electric field modulus $E$, and
therefore for the potential energy of a trapped atom. It is then
simple to derive an expression for the force on the atom along each
direction.

In the case of trap A, due to its symmetries, the potential energy
considered as a power series in $|\vec{r}|$ has a small number of
terms that dominate. The first non-trivial term is a quadrupolar
term, exactly equivalent to (\ref{eq:idealpotquad}):
\begin{equation}
  \mathcal{E}^{\mbox{\tiny{Quad}}} = \alpha ~\frac{3\sqrt{21}}{4\pi}
  \frac{U_1 U_{3}(t) v_{11} v_{22}}{z_0^4}~(2z^2-x^2-y^2).\\
\end{equation}
If, once again, we have $U_3(t)=U_{30} \cos (\omega t)$ then for
each direction the associated Mathieu equation will have a
characteristic parameter $q_i\ (i=x,y,z)$ equal to:
\begin{eqnarray}
q_x=q_y & = & \alpha\frac{3\sqrt{21}}{\pi}
\frac{U_1 U_{30} v_{11}v_{33}}{m_{Rb}\omega^2 z_0^4}, \\
 q_z    & = & -2q_x.
\end{eqnarray}
The second non-trivial term is a term linear in $z$, exactly
equivalent to (\ref{eq:Epotidealgrav}):
\begin{equation}
\mathcal{E}^{\mbox{\tiny{Lin}}}= \alpha ~\frac{\sqrt{15}}{\pi
}\frac{U_1 U_2 v_{11} v_{22}}{z_0^3}~z,
\end{equation}
allowing us to calculate the value of $U_2$ necessary to compensate
gravity.

The use of only these first two terms fits the exact potential
energy to better than 1\% within 100 $\mu$m of $O$ and allows one to
find solutions for the trajectories via analysis of the Mathieu
equation. This treatment, however, does not suffice for trap B due
to its non-zero values of $v_{21}$ and $v_{32}$. Therefore, to treat
trap B, and to achieve a more precise calculation of the
trajectories, we retain a full numerical approach:
\begin{itemize}
\item Each trajectory is computed, using an adaptive Runge-Kutta algorithm.
\item The total time over which we calculate the trajectories
ranges from a few ms to several seconds according to the
application.
\item A trajectory is considered trapped if the atom remains within
the sphere of validity of the numerical fit of the potential (radius
of 400~$\mu$m for trap A, 50~$\mu$m for trap B). 99\% of untrapped
atoms leave this zone within 0.55~s.
\item We use initial conditions for the velocity and position of the
trapped atom coherent with excitation from a trapped cloud of
Rubidium atoms (trapping frequency 1~kHz along each direction) at
thermal equilibrium (temperature $T_0=300$~nK) and with the loading
mechanism that we consider in Sec.~\ref{ssec:loadingmuchip}. These
correspond to Gaussian distributions of widths 5.35 mm/s and
0.27~$\mu$m for the velocity and position respectively. To the
initial velocity is added the single recoil velocity $v_r=6$~mm/s
along the direction $Oy$ that would be received on efficient
adiabatic excitation to the Rydberg state. This temperature and trap
frequency correspond to an atomic cloud close to condensation, or
already condensed (average phonon number $n_{ph}\approx 6$). The
initial conditions that we have taken remain valid however, as long
as the number of atoms in the trap is low enough for mean field
interactions between ground-state atoms to be neglected.
\item For each trajectory we record periodically the atomic position and
velocity, as well as the value of the electric field modulus and its
direction.
\end{itemize}
In all cases we check in the shallow electrodynamic trap that the
trajectory extension, of a much greater size than the initial,
tightly confined ground-state cloud, is also much larger that the
typical de Broglie wavelength of the atom (of the order of
0.50~$\mu$m). This justifies a classical calculation of the motion.

\subsection{Trap performance}\label{ssec:trapperformances}

\begin{figure}
\begin{center}
\resizebox{0.85\columnwidth}{!}{
\includegraphics{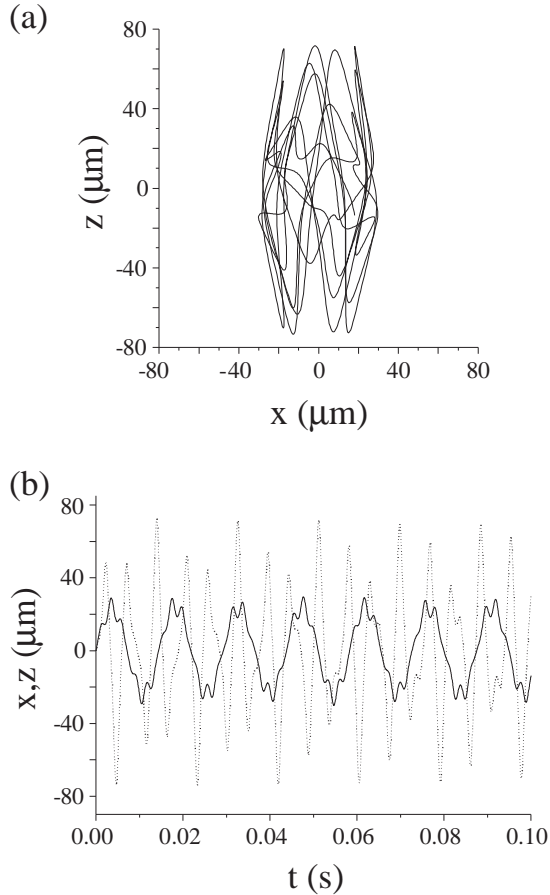}
}
\end{center}
\caption{\label{fig:trajectories} (a) A typical trajectory of an
atom in trap A, shown in the $x0z$ plane. The atom is initially at
the origin of the trap and has a velocity (12,0,12)~mm/s. (b) $x$
(solid line) and $z$ (dotted line) coordinates for the same
trajectory as a function of time. }
\end{figure}

We have studied the trapping characteristics for a wide range of
parameters, especially the voltage $U_1$ which controls the mean
electric field experienced by the atoms, the voltage $U_{30}$ and
the frequency $\omega$ of the a.c. field, the last two together
controlling the confinement. For all of these parameters we find
qualitatively the results predicted by the Mathieu equation. Bound
trajectories are found to be composed of two motions: a fast
micromotion, of frequency $\omega$ and of relatively small
amplitude, and a slower macromotion of larger amplitude and of
frequency $\omega_x = \omega_y \approx \omega_z/2$, an order of
magnitude slower than $\omega$ (see Fig.~\ref{fig:trajectories}). In
the case of non-bound trajectories, the amplitude of the macromotion
usually increases slowly until the atom is finally expelled from the
region in which $V_{\mbox{\tiny{Fit}}}$ is valid.

\begin{figure}
\begin{center}
\resizebox{0.85\columnwidth}{!}{
\includegraphics{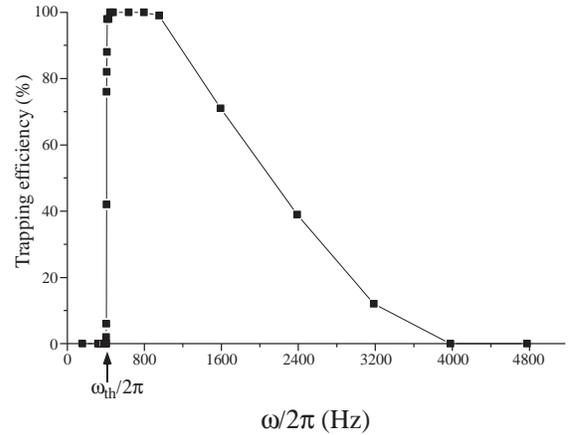}
}
\end{center}
\caption{\label{fig:trappingefficiency} Trapping efficiency (trap A)
as a function of frequency $\omega$. $U_1=$0.2~V, $U_2=-3$~mV,
$U_{30}=$0.056~V, $T_0$=1~$\mu$K.}
\end{figure}

The trapping efficiency as a function of the frequency $\omega$ is
obtained by simulating 100 trajectories and recording the fraction
of them remaining trapped after 1~s. The results for trap A are
shown in Fig.~\ref{fig:trappingefficiency}. No trapping is possible
below a threshold frequency $\omega_{th}/2\pi$. This is in
qualitative agreement with the prediction of the Mathieu equation
stability criteria $|q_z| < 0.907$. For the parameters of
Fig.~\ref{fig:trappingefficiency} this would set a threshold
frequency at 395~Hz, compared to the value of
$\omega_{th}/2\pi=$400~Hz found in the simulations. Just above
$\omega_{th}$ the efficiency depends essentially on the initial
temperature $T_0$ of the sample and can be very close to 100\% for
comparatively cold atoms. If one continues to increase $\omega$ one
observes a slow decay of the efficiency. This can be easily
understood as due to the increasing `averaging' of the
time-dependent electric field seen by the atom. The Mathieu equation
predicts that the macromotion frequencies $\omega_{x}$, $\omega_{y}$
and $\omega_{z}$ decrease as $1/\omega$. There will therefore
rapidly come a point when $\omega$ is so much larger than
$\omega_{x}$, $\omega_{y}$ and $\omega_{z}$ that the time-dependent
field $\vec{E}_3(\vec{r},t)$ averages completely and the atom,
seeing a homogeneous field, is no longer trapped.

In the following, $\omega$ will be chosen slightly above
$\omega_{th}$ where the trapping efficiency is largest. Although
this does not correspond to a situation where a clear cut separation
of the micro- and macromotions exists, we can still distinguish
between the two. We present in Table~\ref{tab:performance} the
frequencies of the macromotion $\omega_\rho = \omega_x = \omega_y$
and $\omega_z$ for different geometries and sets of voltages. In
order to estimate the depth of each trap we calculate the trapping
efficiency for different initial temperatures. We define $T_d$ as
the temperature above which more than 50\% of the atoms are lost
after 1 s. For trap B, two values of $U_1$ are shown (0.2 and
1.5~V), corresponding to two different bias electric field
amplitudes. Comparison of the results clearly show that the smaller
the trap and the larger the bias field the tighter the confinement.
Trap depth can reach as high as the mK with motion frequencies in
the kHz range. However, most of the following simulations have been
performed with a small voltage $U_1$, necessary for studies of
coherent superpositions of $\ket{g}$ and $\ket{e}$, as we shall see
in Sec.~\ref{sec:dephasing}. Even in these unfavourable conditions
for trapping, the depths attained are compatible with typical
cold-atom temperatures and oscillation frequencies remain
reasonable, around 100~Hz.

Small departures of the electric field direction from $\vec{u}_z$ in
the trapping region, due to transverse components of $\vec{E}_2$ and
$\vec{E}_3$, will later prove critical in calculating the efficiency
of the spontaneous emission inhibition
(Sec.~\ref{sec:spontemission}) and the rate of dephasing between
$\ket{g}$ and $\ket{e}$ (Sec.~\ref{sec:dephasing}). We therefore
note in Table \ref{tab:performance} the mean angle $\langle \theta
\rangle$ between the local electric field and $\vec{u}_z$
experienced by an atom over the course of its trajectory.
\begin{table*}
\begin{center}
\begin{tabular}{l|ccccccccc}
\hline\noalign{\smallskip}
 Type & $\eta$ & $U_1$ (V) & $U_{30}$ (V) & $U_2$ (mV) & $\omega/2\pi$ (Hz)
& $\omega_{\rho}/2\pi$ (Hz) & $\omega_z/2\pi$ (Hz) & $T_d$ ($\mu$K) & $\langle
\theta \rangle$ (mRad) \\
\noalign{\smallskip}\hline\noalign{\smallskip}
 Trap $A$ & 4.49 & 0.2 & 0.056 & -3 & 430 & 64 & 175 & 180 & 9.38 \\
 Trap $B$-$\alpha$ & 0.05 & 1.5 & 0.5 & 0 & 20700 & 1460 & 2910 & 1000 &
4.22 \\
 Trap $B$-$\beta$ & 0.05 & 0.2 & 0.14 & -0.45 & 2860 & 207 & 414 & 35 &
3.57 \\
\noalign{\smallskip}\hline
\end{tabular}
\end{center}
\caption{\label{tab:performance}Trap $A$ and trap $B$ correspond to
the geometries of Fig.~\ref{fig:TrapGeometry}(b) and (c)
respectively. Labels $\alpha$ and $\beta$ for Trap $B$ correspond to
different voltage settings. For each trap, we give the values for
$U_1$, $U_2$, $U_{30}$ and the frequency $\omega/2\pi$. The
frequencies $\omega_z/2\pi$ and $\omega_{\rho}/2\pi$ are the typical
longitudinal (along $Oz$) and transverse oscillation frequencies in
the trap. The trap depth $T_d$ is defined as the temperature for
which half of the atoms remain trapped. $\langle \theta \rangle$ is
the mean angle between the local electric field $\vec{E}$ and
$\vec{u}_z$, averaged over 100 trajectories at temperature $T_d/2$.}
\end{table*}

\section{Loading of the trap, preparation of a subpoissonian
sample of Rydberg atoms}\label{sec:loading}

\subsection{The dipole-blockade effect}\label{ssec:dipoleblockade}

For many quantum information processing experiments, the
deterministic preparation and storage of an individual qubit is a
critical requirement. This has already been achieved, for example,
with trapped ions. In previous experiments with circular Ryd\-berg
atoms \cite{ENS_RMP}, we expressly set the rate of excitation
extremely low in order to be able to neglect two-atom events. This
is, of course, at the expense of very long data acquisition times.
We propose here to deterministically load the trap with a single
Rydberg atom by using the dipole-blockade effect. This phenomenon
has been proposed as an efficient way to perform quantum gates
between two Rydberg atoms~\cite{QI_LUKINDIPOLEBLOCKADE01} but can
also ensure the preparation of one and only one low angular momentum
Rydberg state by laser excitation of a dense micron-sized cloud of
ground-state atoms~\cite{TR_SAFFMANN02}. The essential idea behind
this phenomenon is that level shifts due to the dipole-dipole
interaction make the laser nonresonant for the excitation of more
than one Rydberg atom.

\begin{figure}
\begin{center}
\resizebox{0.85\columnwidth}{!}{
\includegraphics{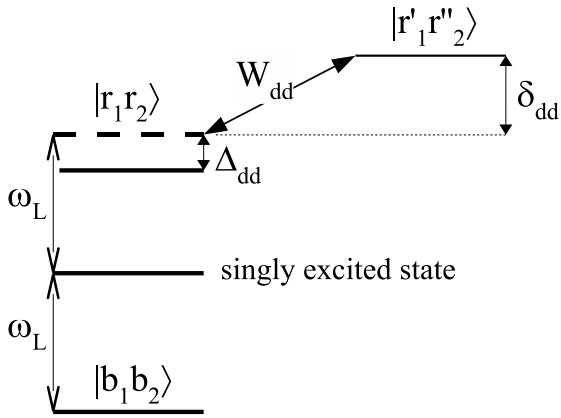}
}
\end{center}
\caption{\label{fig:levelsdipblockade} Level diagram demonstrating
the dipole-blockade effect.}
\end{figure}

In order to assess the effect of atom-atom interactions we consider
an ensemble of $N$ cold ground-state atoms (temperature in the
$\mu$K range) in the presence of a laser resonant with the
transition between a low-lying state $\ket{b}$ and a given Rydberg
state $\ket{r}$. Atomic motion can be neglected during the laser
excitation and hence each atom position $\vec{R}_i$ is treated as
fixed. Let us assume that the initial trap confines the atoms in a
micron-sized region such that $ |\vec{R}_{ij}|=|\vec{R}_i-\vec{R}_j|
< 1\ \mu$m for any couple $(i,j)$ of atoms. We then switch on the
laser light, of frequency $\omega_L$.

Let us consider for a moment only atoms 1 and 2.
Fig.~\ref{fig:levelsdipblockade} shows the relevant levels for this
simplified situation. We tune the laser frequency to be resonant on
the transition from the lower state $\ket{b_1 b_2}$ to the singly
excited state, where one of the atoms has been excited to the
Rydberg state $\ket{r}$. In the absence of interactions, this laser
would also be resonant with the transition towards the state
$\ket{r_1 r_2}$ where both atoms 1 and 2 are excited into the state
$\ket{r}$. We must, however, take into account the coupling of this
level $\ket{r_1 r_2}$ with other doubly excited states of different
Rydberg levels, for example $\ket{r_1^\prime r_2^{\prime \prime}}$,
shifted in energy from $\ket{r_1 r_2}$ by $\delta_{dd}$ (see
Fig.~\ref{fig:levelsdipblockade}). This coupling, of magnitude
$W_{dd}$, will cause a shift $\Delta_{dd}$ of the level $\ket{r_1
r_2}$ which, if large enough, will make the excitation of the doubly
excited level nonresonant and therefore strongly suppressed. We must
therefore calculate $\Delta_{dd}$. First order perturbation theory
gives us:
\begin{equation}
\Delta_{dd}=\frac{W_{dd}^2}{\delta_{dd}}.
\end{equation}
$\delta_{dd}$ will be of the order of the separation of Rydberg
levels, a few GHz for $n\sim50$. A classical order of magnitude for
$W_{dd}$ is given by:
\begin{equation}
h W_{\mbox{\tiny{dd}}}=\frac{n^4 e^2 a_0^2}{8\pi \epsilon_0
|\vec{R}_{12}|^3},
\end{equation}
where $a_0$ is the Bohr radius. For a distance $|\vec{R}_{12}|=1\
\mu$m and $n=50$ one calculates $W_{\mbox{\tiny{dd}}}=3$~GHz. We
therefore calculate a shift of the doubly excited level
$\Delta_{dd}\sim$~1~GHz. This order of magnitude calculation is in
agreement with the result $\Delta_{dd}=100$~MHz found by Saffmann
and Wal\-ker~\cite{TR_SAFFMANN02}. This shift is easily resolved
using modern lasers and hence excitation of multiple Rydberg atoms
will be out of resonance with the excitation laser. It is shown in
\cite{TR_SAFFMANN02} that for the resonant excitation of the
transition $\ket{b}\longrightarrow\ket{r}$ in a cloud of ground
state atoms of diametre 5~$\mu$m using a laser of Rabi frequency
1~MHz one obtains multiple- or non-excitation of $\ket{r}$ for less
than 1 event in $10^4$ (for a number of ground-state atoms of the
order of a few hundred, compatible with the assumption that we can
neglect interactions in the calculation of the initial conditions).

\subsection{Loading from a magnetic atom-chip trap}\label{ssec:loadingmuchip}

Atom-chip traps~\cite{MX_HANSCHCHIP99} can fulfill all the
conditions required by our proposal. They allow for very high
densities of ground-state atoms in trap volumes as small as a few
$\mu$m$^3$~\cite{FOLMANCHIPREV_02}. They can, by construction, store
atoms close to surfaces and the operation of a conveyor belt, ideal
for bringing atoms from a capture region into the Rydberg trap, has
been proven~\cite{TR_CONVEYOR_REICHEL01}. Finally, evaporative
cooling in such traps allows one to attain temperatures as low as a
few hundred nK~\cite{MX_HANSCHBECCHIP01}. The need for operation at
cryogenic temperatures, in order to get rid of blackbody radiation
which would rapidly destabilize the Rydberg states (see
Sec.~\ref{sec:spontemission}), makes it impossible however to use
standard atom-chip designs with normal conductors. We plan to use
instead superconductors to create the magnetic fields. This would
allow us to pass current without Joule heating as long as one
remains below the critical current. In standard atom-chip
experiments at room temperature the current densities are typically
of the order of $5 \times 10^6$~Acm$^{-2}$
\cite{SCHMIEDMAYERCHIPFAB_04}. We have checked that it is possible
to reach similar current densities with our superconducting wires.
In the case of a superconducting Niobium wire of thickness 1~$\mu$m
and width 10~$\mu$m, sputtered on a Silicon Oxide substrate, we find
$I_c=0.5$~A, corresponding to a current density of $5 \times
10^6$~Acm$^{-2}$.

Thanks to our microfabricated design, it is simple to integrate our
Rydberg trap with the magnetic atom-chip trap and also the
electrodes necessary for the transfer of the atom from a low angular
momentum Rydberg state to a circular state \cite{ENS_CIRCRB}. We
have designed a multi-layer configuration with the electrodes
necessary for the Rydberg trap placed on top of the wires of the
magnetic atom chip. The circularization is achieved using wires on
the magnetic atom-chip layer to create the necessary magnetic field
along with the application of a r.f. voltage to electrodes on the
Rydberg trap layer. The Rydberg trap electrodes will not
significantly affect the magnetic field created by the wires of the
lower layer as long as a normal conductor such as Gold is used.
Evaporative cooling in the magnetic atom-chip trap can provide a few
hundred atoms in a micron-sized cloud at temperatures as low as a
few hundred nK~\cite{MX_HANSCHBECCHIP01}, conditions ideal for the
operation of the dipole-blockade effect. The magnetic trap is then
suddenly switched off and the excitation towards a low angular
momentum Rydberg state performed immediately. Carefully designed
laser excitation schemes result in the atom receiving only one
optical photon recoil, just before the circularization process,
itself lasting about 20~$\mu$s~\cite{ENS_CIRCRB,ENS_RMP}. The
flexibility of magnetic atom chips will allow us to superimpose the
centre of the magnetic trap with that of the Rydberg trap.

\section{Making the atom long-lived by spontaneous emission
inhibition}\label{sec:spontemission}

\subsection{Principles}\label{ssec:spontinhibprincip}

\begin{figure}
\begin{center}
\resizebox{0.85\columnwidth}{!}{
\includegraphics{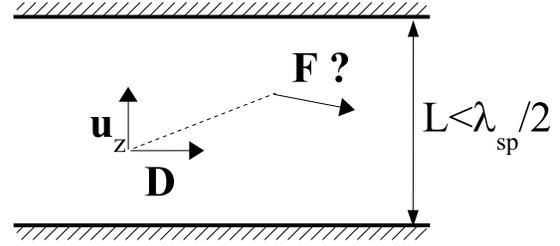}
}
\end{center}
\caption{Ideal geometry for spontaneous emission
inhibition.\label{fig:inhibTheorie}}
\end{figure}

The basic idea behind the inhibition of spontaneous emission is to
place the atom inside a cavity containing no mode into which the
atom can emit its photon upon decay~\cite{ENS_HOUCHES90}. For the
sake of simplicity we first of all consider the geometry of
Fig.~\ref{fig:inhibTheorie} where a circular Rydberg atom is placed
between two infinite, perfectly conducting planes separated by a
distance $L$. This situation is relatively close to that of
Fig.~\ref{fig:TrapGeometry}~(b). We now assume that there exists
between the two planes a d.c. electric field
$\vec{E}_{\mbox{\tiny{dir}}}$ orthogonal to the planes (and hence
parallel to their normal $\vec{u}_z$). Obeying the standard
selection rules, a circular Rydberg state of principal quantum
number $n$ possesses only a single possible decay channel, this
being towards the lower circular state $n-1$ (for example $\ket{e}
\rightarrow \ket{g}$). In the process it emits a photon of
polarization $\sigma^{+}$ with respect to $\vec{u}_z$, of wavelength
$\lambda_{sp}$ ($\lambda_{sp}\simeq6$~mm for the decay of $\ket{g}$
and $\ket{e}$). Due to its polarization, the electric field
associated with this photon is perpendicular to $\vec{u}_z$ and
hence parallel to the planes. The maximum wavelength of any such
mode inside this cavity is $2L$ as the electric field must cancel at
the surface of each plane. Therefore, if $\lambda_{sp}>2L$, there
exists no cavity mode at resonance with the spontaneous emission
transition and the atom cannot decay, remaining in the circular
state for an infinitely long time.

We now detail a more rigorous, semi-classical approach to this
phenomenon, as found in Ref.~\cite{ENS_HOUCHES90}. We calculate the
effect of the field radiated by the dipole upon itself, and hence
the consequent energy shifts and lifetimes. The electric field
radiated by a dipole $\vec{D}=D\exp (-i\omega t)\vec{u}_d$ at
position $\vec{r}_d$ oscillating at frequency $\omega$ along unit
vector $\vec{u}_d$ is given by:
\begin{equation}
\vec{E}_{dip}(\vec{r},t)=\frac{D\omega^3}{4\pi\epsilon_0c^3}
\bm{\mathcal{F}}^{(cav)}(\vec{r},\vec{r}_d, \omega)e^{-i\omega t},
\end{equation}
where $\bm{\mathcal{F}}^{(cav)}(\vec{r},\vec{r}_d,\omega)$ is the
susceptibility of the field inside the cavity, characterizing the
linear response of the system to the dipole movement. This
susceptibility can be written in the form:
\begin{equation}
\bm{\mathcal{F}}^{(cav)}(\vec{r},\vec{r}_d,\omega) =
\bm{\mathcal{F}}^{(0)}(\vec{r},\vec{r}_d,\omega) +
\bm{\mathcal{F}}^{(r)}(\vec{r},\vec{r}_d,\omega).
\end{equation}
The field corresponding to
$\bm{\mathcal{F}}^{(0)}(\vec{r},\vec{r}_d,\omega)$ is the field
radiated by the dipole in free space and that corresponding to
$\bm{\mathcal{F}}^{(r)}(\vec{r},\vec{r}_d,\omega)$ reflects the
modification introduced by the presence of the cavity and is easily
understood as being the field radiated by the images of the dipole
in the surfaces of the cavity. When we calculate the interaction of
the dipole with the total field we will therefore have two terms:
\begin{itemize}
\item A first representing the interaction of the dipole with its
own free space field. This term leads us to the calculation of the
natural lifetime of the dipole and the Lamb shift of its frequency.
This calculation however is plagued by divergences. Their values
must be derived by a fully quantum treatment, and we simply insert
these results into our calculation.
\item A second corresponding to the interaction of the dipole with
the field of its images. This leads to the modification of the
natural lifetime (that we search to calculate) along with the
accompanying frequency shift. Its calculation, in contrast to the
first, poses no problem.
\end{itemize}

\begin{figure}
\begin{center}
\resizebox{0.85\columnwidth}{!}{
\includegraphics{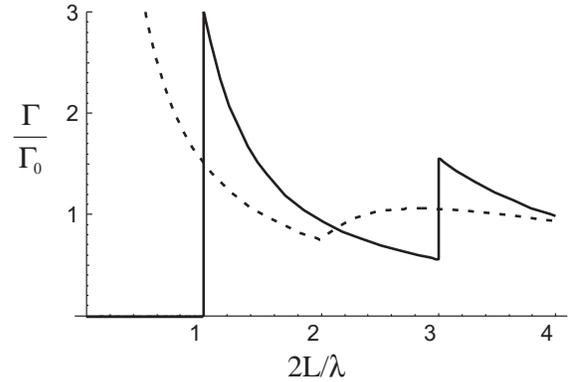}
}
\end{center}
\caption{Spontaneous emission rate $\Gamma$, in units of
$\Gamma_{0}$, as a function of the cavity plane separation $L$. The
dipole is either parallel to the mirrors ($\Gamma_{\parallel}$,
continuous line) or perpendicular to them ($\Gamma_{\perp}$, dashed
line). Both are calculated for the case of perfectly conducting
mirrors. \label{fig:InhibMiroirParfait}}
\end{figure}

This approach gives the following expression for the rate of decay
of the dipole (atom) inside the cavity~\cite{ENS_HOUCHES90}:
\begin{equation}
\Gamma=\Gamma_{0}\left (1+\frac{3}{2}\mbox{Im}\left
[\vec{u}_d.\bm{\mathcal{F}}^{(r)}(\vec{r} _d,\vec{r}_d,\omega)\right
]\right ),
\end{equation}
the imaginary part of the susceptibility corresponding to the
component of the image field in quadrature with the dipole
oscillation and hence responsible for dissipation. We present in
Fig.~\ref{fig:InhibMiroirParfait} the evolution of the ratio $\Gamma
/ \Gamma_{0}$ as a function of the distance $L$ in two specific
situations: $\vec{D}$ parallel to $\vec{u}_z$; $\vec{D}$
perpendicular to $\vec{u}_z$. In the latter, the case of $\sigma$
emission in the presence of a quantization axis along $\vec{u}_z$,
one observes a complete inhibition for $L<\lambda /2$. In the
former, the case of $\pi$ emission in the presence of the same
quantization axis along $\vec{u}_z$, the dipole couples to modes
obeying different boundary conditions and its emission is exalted
for small $L$ values. In the case of a $\sigma$ transition in the
presence of a quantization axis slightly tilted with respect to
$\vec{u}_z$ we have components of $\vec{D}$ both parallel and
perpendicular to $\vec{u}_z$ and consequently the possibility of
decay via both inhibited and exalted channels.
Inhibition has already been observed in the microwave
domain~\cite{QC_KLEPPNERINHIBITION85} as well as for optical
transitions~\cite{QC_YALEOPTICALINHIB87}. A value of less than 0.1
for $\Gamma/\Gamma_0$ has been reported in the case of Rydberg
atoms. This would already increase the radiative lifetime of
circular states to a significant fraction of a second.

\subsection{Limitations of the inhibition}\label{ssec:spontinhibimperf}

\begin{figure}
\begin{center}
\resizebox{0.85\columnwidth}{!}{
\includegraphics{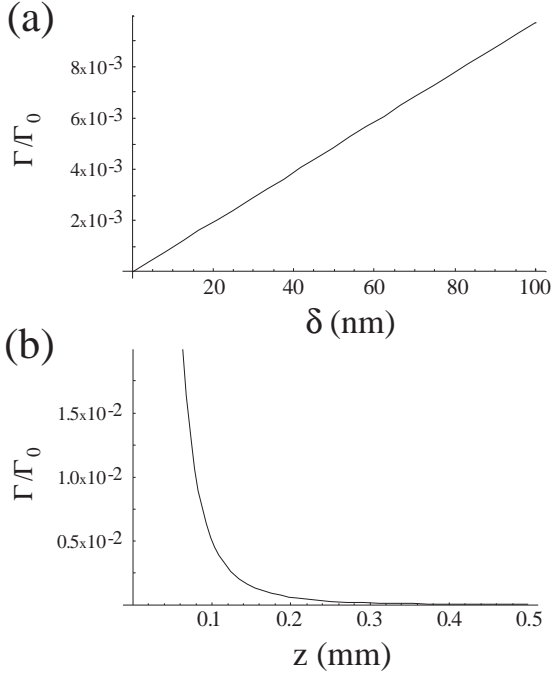}
}
\end{center}
  \caption{\label{fig:InhibEpaisseurdePeau} (a) Spontaneous emission inhibition
factor $\Gamma_{\parallel}/\Gamma_{0}$ as a function of the skin
depth $\delta$. The conditions are the same as in
Fig.~\ref{fig:inhibTheorie} with $L=1$~mm, $\lambda=6$~mm and a
dipole 120~$\mu$m away from one mirror. (b)
$\Gamma_{\parallel}/\Gamma_{0}$ as a function of the atom-mirror
separation for the same values of $L$ and $\lambda$ as (a) and with
$\delta=30$~nm, corresponding to the case of Gold at 1~K }
\end{figure}

The ideal case presented in Sec.~\ref{ssec:spontinhibprincip} is,
however, far from the realistic geometries of
Fig.~\ref{fig:TrapGeometry}(b) and (c), the imperfections of which
we must now evaluate. As a first example, the electrodes do not form
perfect infinite mirrors. It is possible to neglect the effect of
their finite size if, as will be the case, the transverse dimensions
of the trap are much larger than $\lambda_{sp}$. A more critical
effect comes from the residual absorption by the imperfect mirrors,
dissipating the energy radiated by the atom and limiting its
lifetime. This effect must therefore be included in the calculation
of the susceptibility
$\bm{\mathcal{F}}^{(r)}(\vec{r},\vec{r}_d,\omega)$. The dipole of an
image produced by a number of reflections in the mirrors $N_{R}$ is
therefore multiplied by a factor:
\begin{equation}
(\rho e^{i\chi})^{N_{R}},
\end{equation}
where $|\rho|^2$ is the reflection coefficient of the mirror and
$\chi$ accounts for the phase shift at each reflection. For trapping
electrodes made of ordinary conductors (such as Gold) $\rho$ and
$\chi$ are related to the skin depth $\delta$ by the following
expressions:
\begin{eqnarray}
\rho & \simeq & \exp(-\frac{2 \pi \delta}{\lambda}), \\
\chi & \simeq & \frac{2 \pi \delta}{\lambda},
\end{eqnarray}
provided that, as is the case here, $\delta \ll
\lambda$~\cite{TXT_JACKSON}.
Figure~\ref{fig:InhibEpaisseurdePeau}(a) presents the inhibition
fact\-or $\Gamma_{\parallel}/\Gamma_{0}$ as a function of the skin
depth $\delta$ in the case of a dipole radiating at 50~GHz
($\lambda=6$~mm). The cavity spacing $L$ is 1~mm and the dipole
120~$\mu$m away from one mirror as in trap B. One can see that
inhibition is still efficient (a factor of greater than 100) for
skin depths of less than 100~nm. Operating, as we must, under
cryogenic conditions, we can therefore use Gold, for which $\delta
=30$~nm at 1~K. For imperfect mirrors there is another effect that
must be considered. The inhibition factor now depends upon the
position of the atom between the two planes.
Figure~\ref{fig:InhibEpaisseurdePeau}(b) shows this effect. The
closer to the surfaces, the worse the inhibition. One might fear
that this would pose problems in trap B. However, even in the case
of a circular Rydberg atom in trap B, 120~$\mu$m away from Gold
trapping electrodes at 1K, the inhibition factor of 340 would imply
a radiative lifetime of 10~s.

We must also consider the consequences of our quantization axis,
following adiabatically the electric field $\vec{E}(\vec{r},t)$,
being not always perfectly perpendicular to the plane mirrors. A
non-zero angle $\theta(\vec{r},t)$ between $\vec{E}(\vec{r},t)$ and
$\vec{u}_z$ will be responsible for a small component $D_\perp = D
\sin \theta$ of the dipole being orthogonal to the mirrors (and for
a reduced dipole $D_\parallel = D \cos \theta$ parallel to the
mirrors). It is therefore possible for the atom to couple to modes
possessing exalted decay. Taking into account only this decay
channel, the probability for remaining in the excited state is
therefore given by:
\begin{eqnarray}
P & = & \exp \left( -\int_{\mbox{\tiny tr.}} \Gamma_{\perp} \sin ^2
\theta\left[\vec{r}(t),t\right] dt \right ) \nonumber \\
 & = & \exp \left( - \Gamma_0 \int_{\mbox{\tiny tr.}}
\frac{\Gamma_{\perp}}{\Gamma_0} \sin ^2
\theta\left[\vec{r}(t),t\right] dt \right ),
\label{eq:inhibcorrectedangle}
\end{eqnarray}
where the integral is performed along the trajectory followed by the
atom.

As we have seen (Table~\ref{tab:performance}), $\theta \ll 1$ for
the trajectories considered and we can therefore approximate
$D_\parallel = D$ and $\sin^2 \theta = \theta ^2$. From
Eq.~\ref{eq:inhibcorrectedangle} one can deduce a corrected
inhibition factor in $\Gamma_{\mbox{\tiny corr.}}/\Gamma_0$, related
to the value of $\theta$ along a trajectory:
\begin{equation}
\frac{\Gamma_{\mbox{\tiny corr.}}}{\Gamma_0}=
\frac{\Gamma_{\parallel}}{\Gamma_0}+
\frac{\Gamma_{\perp}}{\Gamma_0}\overline{\theta^2},
\end{equation}
where:
\begin{equation}
\overline{\theta^2}=\frac{1}{t_{tr.}}\int_{\mbox{\tiny tr.}} \theta
^2 dt,
\end{equation}
is the average of the square of the angle $\theta$ over the total
duration of the trajectory, $t_{tr.}$. For $L=1$~mm, $\lambda=6$~mm
and $\delta = 30$~nm we find an exaltation factor
$\Gamma_{\perp}/\Gamma_0$ of 4.5, whatever the distance between the
dipole and the mirrors. Under the conditions considered
(Table~\ref{tab:performance}), $\Gamma_{\mbox{\tiny
corr.}}/\Gamma_0$ remains greater than 150.

The final effect that must be considered in evaluating the residual
lifetime of a circular Rydberg atom in the trap is excitation by
blackbody photons in the modes of the cavity. These can induce
$\pi$-polarized transitions between $\ket{g}$ (resp. $\ket{e}$) and
the other states of the $m=49$ (resp. $m=50$) multiplicity (for
example $\ket{g} \rightarrow \ket{i}$, see Fig.~\ref{fig:Stark}).
The rate of this process is proportional to the mean photon number
$n_{t}$ at the relevant transition frequency times the relaxation
rate $\Gamma_\pi$ for each transition, the total rate
$\Gamma_{\mbox{\tiny{B.B.}}}$ being equal to the sum of all possible
transitions $\Gamma_{\mbox{\tiny{B.B.}}}(k)=\sum_{k^\prime}n_t(k
k^\prime)\Gamma_\pi (k k^\prime)$. We calculate
$\Gamma_{\mbox{\tiny{B.B.}}}(g)=3.15~\mbox{s}^{-1}$ and
$\Gamma_{\mbox{\tiny{B.B.}}}(e)=2.75~\mbox{s}^{-1}$, the effect of
exaltation of $\pi$ transitions having been taken into account. If
the trap is cooled down to 1~K, one has $n_t=0.1$ and a
corresponding radiative lifetime of about 3~s, once again not a
limiting factor.

\section{Preserving atomic coherences over long times}\label{sec:dephasing}

To this point, we have shown that it is possible to trap a single
atom in a circular Rydberg state over times in the second range. It
is, however, important to see if one can manipulate and maintain its
internal state with the same precision. More precisely, if we hope
to use this technique in quantum information experiments or for high
precision spectroscopy, it is important to verify that one can
prepare an atom in a coherent superposition of states and probe its
phase at a later time. In the following we will focus on the
transition between the two circular states $\ket{e}$ and $\ket{g}$.

\subsection{Electric shifts as the main cause of
dephasing}\label{ssec:electricshifts}

As we saw in Sec.~\ref{ssec:cra}, the Stark polarizabilities of
levels $\ket{e}$ and $\ket{g}$ are slightly different.
Fig.~\ref{fig:compensation} shows the energy levels as a function of
the electric field amplitude. The frequency $\nu_{eg}=(\mathcal{E}_e
- \mathcal{E}_g)/h$ of the transition has a strong dependence on the
electric field:
\begin{equation}
\delta\nu_{eg} =\nu_{eg}(E)-\nu_{eg}(0)= \delta\alpha_{eg}E^2,
\end{equation}
with $\delta\alpha_{eg}=-25.5\mbox{~Hz/(V/m)}^2$. If one considers
the case of an atom in trap $A$ in the conditions of
Table~\ref{tab:performance}, prepared from a cloud at 0.3 $\mu$K, it
experiences over its trajectory a mean electric field amplitude
$E_a=400$ V/m, with an excursion of $\Delta E =\pm 1$~V/m. The
corresponding frequency broadening:
\begin{equation}
\Delta \nu_{eg}=2 \delta\alpha_{eg}E_a \Delta E,
\end{equation}
is of the order of 20~kHz. This broadening is inhomogeneous as it is
different from one trajectory to the next and could only be
controlled by perfect control of the initial conditions of the
atomic motion. The dephasing time associated to this motional
broadening is of the order of a few tens of $\mu$s.

In addition to this dephasing, the trapping frequencies
($\omega_\rho$, $\omega_z$) for states $\ket{e}$ and $\ket{g}$
differ by about 10\%. The trajectories for the two states are
therefore rapidly separated (`Stern Gerlach' effect). Coherence
would be lost when this separation exceeds the wave-packet coherence
length, of the order of the de Broglie wavelength (about 0.5~$\mu$m
for the conditions considered).

\subsection{Tailoring the atomic levels to cancel
electric shifts}\label{ssec:leveldressing}

A similar situation is observed for ground-state atoms in magnetic
or optical dipole traps. The potential of these traps is also, in
general, level dependent. However, well chosen trap laser
wavelengths~\cite{TR_KATORI03} or bias magnetic
fields~\cite{TR_CORNELLCOHERENCE02,TR_REICHELCOHERENCE04} minimize
the effect. To achieve the same end in the Rydberg trap we propose
to use a microwave field dressing in order to tailor the atomic
energies. A $\pi$-po\-la\-ri\-zed microwave field, of frequency
$\omega_0$, is fed inside the trap. This field is compatible with
the trap boundary conditions. For the sake of clarity, and as
alluded to earlier, we consider that additional boundary conditions,
far from the trap center, support a propagating wave in a guided
mode: its amplitude, independent of $z$ and $x$, is maximal at the
origin and varies sinusoidally with $y$, canceling at $y=\pm 1$~ cm.

\begin{figure}
\begin{center}
\resizebox{0.85\columnwidth}{!}{
\includegraphics{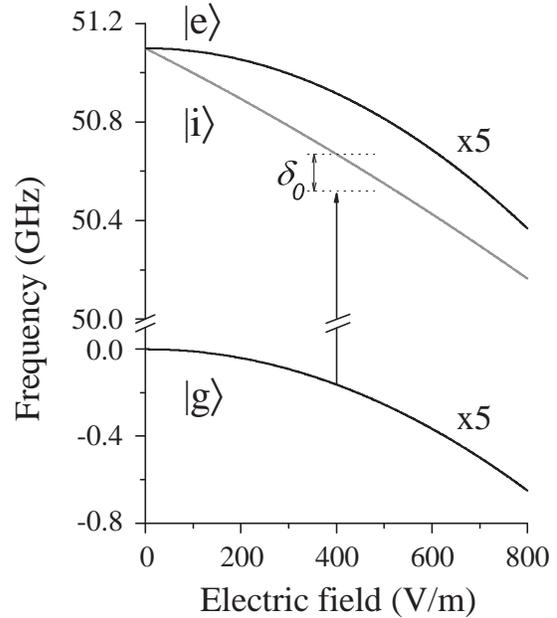}
}
\end{center}
\caption{\label{fig:compensation} Energies of levels $\ket{g}$,
$\ket{e}$ and $\ket{i}$ as a function of the electric field. States
$\ket{g}$ and $\ket{e}$ have a small, approximately quadratic Stark
effect, magnified in this figure by a factor 5 for the sake of
clarity. Level $\ket{i}$ has a much larger, linear Stark effect. The
microwave dressing (vertical arrow) predominantly mixes $\ket{g}$
and $\ket{i}$ and reduces the difference in Stark polarizability
between $\ket{e}$ and $\ket{\widetilde{g}}$, the resulting dressed
level.}
\end{figure}

One can easily understand the effect of the microwave by the
consideration of a simple two-level approximation where the
microwave only couples levels $\ket{g}$ and $\ket{i}$ (see
Fig.~\ref{fig:Stark}). The state $\ket{i}$ experiences a large,
linear Stark effect [1~MHz/(V/m), see Fig.~\ref{fig:compensation}].
We define $\Omega_0$ as the classical Rabi frequency on this
particular transition, dependent on the applied microwave power. The
field is detuned by $\delta_0=\omega_{gi}(E_a)- \omega_0>0$ to the
red of the $\ket{g}/\ket{i}$ transition, Stark shifted by the
average field $E_a$. As a result the `dressed' level
$\ket{\widetilde{g}}$ acquires a fraction of the Stark
polarizability of $\ket{i}$ and the dependence of its energy on
electric field can be made parallel to that of $\ket{e}$, at least
in the neighbourhood of $E_a$. More precisely, we can choose the
values of the two independent parameters $\delta_0$ and $\Omega_0$
so as to cancel the linear and quadratic terms in the expansion of
the dressed state transition frequency $\omega_{e\tilde{g}}(E)$
around $E_a$. For $E_a=400$~V/m, the resulting $\delta_0=2\pi \times
746.158$~MHz and $\Omega_0=2\pi \times 228.442$~MHz have reasonable
values. The remaining higher order terms in the expansion of the
transition frequency $\omega_{e\tilde{g}}(E)$ are smaller than
0.05~Hz over the complete field amplitude range of $\Delta E = \pm
1$~V/m around $E_a$. This would result in dephasing times longer
than 20~s.

We note here that this scheme would also compensate for the problem
of patch effects on the surface of the electrodes if the field which
they create remains relatively small compared to the directing field
amplitude $E_a$. This effect is caused by the misalignment of
crystal orientation in adjacent crystals in the metal. Let us
consider a surface with voltage difference $\Delta V$ between
adjacent crystals of characteristic size $a$. An ensemble of such
patches will create at a distance $d$ above the surface a field:
\begin{equation}
E_{\mbox{\tiny patch}} = \frac{a \Delta V}{d^2}.
\end{equation}
Over a volume of dimension $\Delta r$, this field has an
inhomogeneity of the order of:
\begin{equation}
\Delta E_{\mbox{\tiny patch}} = \frac{a \Delta V \Delta r}{d^3}.
\end{equation}
Patch-effects for Gold electrodes were measured at much closer
distances in Ref.~\cite{QC_YALESHIFTS96} where they found
$E_{\mbox{\tiny patch}} \simeq 10^4$~V/m for $d=0.25\ \mu$m and $a
\simeq 30$~nm. Being conservative, we assume the patch size to be
100~nm. Building on these figures we estimate the field to be
$0.008\pm 0.0003$~V/m for trap A and $0.15\pm 0.005$~V/m for trap B,
the inhomogeneity being calculated over a volume of dimension
$\Delta r = 15\ \mu$m and 5~$\mu$m, the typical excursion of atoms
in traps A and B respectively. All of these values remain much
smaller than $E_a$ and would hence be well compensated for by our
technique.

\subsection{Assessment of performances}\label{ssec:dressingimperf}

The simple explanation of Sec.~\ref{ssec:leveldressing} is, however,
far from describing the true physical reality:
\begin{itemize}
\item The value calculated for $\delta_0$ is not significantly smaller
than the energy difference between state $\ket{i}$ and state
$\ket{i'}$ (see Fig.~\ref{fig:Stark}), which is therefore also
significantly coupled to $\ket{g}$. Moreover the microwave field
also couples state $\ket{e}$ to the $n=52$ manifold. It is therefore
important to adopt a multilevel approach in order to account for the
coupling by the $\pi$-polarized dressing microwave of all the levels
of equal $m$ (see Fig.~\ref{fig:Stark}).
\item This coupling therefore mixes the level $\ket{g}$ with other
levels of the same $m=49$ (as is its aim) and equally $\ket{e}$ with
levels of $m=50$. It must not be forgotten, however, that these
levels have finite, potentially short, lifetimes in the cavity. We
can therefore now see a form of spontaneous emission by
$\ket{\widetilde{g}}$ and $\ket{\widetilde{e}}$ in which they absorb
a photon from the dressing microwave mode and immediately re-emit it
into another mode of the cavity. These spontaneous events would
destroy any coherence existing between $\ket{\widetilde{g}}$ and
$\ket{\widetilde{e}}$. We must therefore quantify the rate at which
these events occur and check that the corresponding lifetime is not
limiting.
\item The dressed level transition frequency $\omega_{e\tilde{g}}$ is
dramatically dependent on the microwave Rabi frequency, previously
considered constant and equal to $\Omega_0$. But $\Omega$ is
actually time- and position-dependent due to the spatial profile of
the mode (nodes at $y=\pm1$~cm, see Sec.~\ref{ssec:leveldressing})
and due to the variation of $\theta(\vec{r},t)$. The instantaneous
Rabi frequency experienced by an atom is therefore given by
$\Omega(\vec{r},t)=\Omega_0 f(\vec{r}) \cos \theta(\vec{r},t)$. This
variation of the Rabi frequency creates an inhomogeneous broadening
which must be included in our final calculation.
\item The tilt $\theta$ also couples a fraction of the microwave power to
$\sigma$ transitions within the Rydberg levels, with an effective
Rabi frequency $\Omega_\sigma (\vec{r},t)\propto\Omega_0 f(\vec{r})
\sin \theta(\vec{r},t)$, significantly smaller than $\Omega$. The
large detuning $\delta_0$ prevents us from having resonant
one-photon transitions between the levels $\ket{e}$ and $\ket{g}$
due to this coupling. Nevertheless, there could exist a multiphoton
transition coupling the circular states to adjacent manifolds. A
1~Hz Rabi frequency would be sufficient to transfer population from
the circular state to this adjacent one, and limit the coherence
time between $\ket{e}$ and $\ket{\widetilde{g}}$ to 1~s. We must
therefore check that no such resonance exists.
\item To achieve the necessary precision in the position of the levels
$\ket{e}$, $\ket{g}$, $\ket{i}$ and $\ket{i^\prime}$ we must go
beyond the quadratic Stark expansion of Eq.~\ref{eq:stark0-2}.
\end{itemize}

We will now address these points one by one, turning first of all to
the last. To achieve the precision necessary we diagonalize the
Stark Hamiltonian of the circular state and the 4 manifolds of
greater $n$ above it, in the presence of an electric field. The
resulting energies were fitted by a 4th order polynomial over
electric fields from 0 to 1000~V/m with an error less than 0.5~Hz.
It was checked that adding a 5th manifold above the circular state
to the Hamiltonian changed the resulting energies by less than 1~Hz.

In order to take account of the coupling of levels other than $\ket{g}$
and $\ket{i}$ by the microwave field, we must carry out a diagonalization
of a Hamiltonian containing all relevant couplings. The microwave
field is fully defined by its classical Rabi frequency at trap
centre $\Omega_0$ on the $\ket{g} \rightarrow \ket{i}$ transition
and its detuning $\delta_0$. The angle $\theta (\vec{r},t)$ being
small, we consider firstly only the $\pi$-polarized component of the
microwave field, coupling levels of equal $m$. We have adopted a
dressed level approach to this problem and therefore consider the
ladder of levels $\ket{n,n_1,m,n_\pi}$ where the additional quantum
number $n_\pi$ counts the number of photons in the microwave field.
For the case $m=49$, relevant for level $\ket{g}$, the coupling
between initial and final states $\ket{n_i,n_{1,i},49,n_\pi+1}$ and
$\ket{n_f,n_{1,f},49,n_\pi}$ is given by:
\begin{equation}
\Omega_{n_i,n_{1,i}}^{n_f,n_{1,f}}(n_\pi)=\Upsilon_{n_i,n_{1,i}}^{n_f,n_{1
,f}} \sqrt{n_\pi +1}
\end{equation}
where $\Upsilon_{n_i,n_{1,i}}^{n_f,n_{1,f}}$  is the Rabi frequency
on the considered transition for $n_\pi =0$, proportional to the
dipole matrix element $\mathcal{M}_{n_i,n_{1,i}}^{n_f,n_{1 ,f}}$.
However, as the dressing microwave is a classical field (and hence
$n_\pi \gg 1$), one can assume $\sqrt{n_\pi}\approx
\sqrt{n_\pi+1}\approx \mbox{constant}$. Therefore each coupling
strength can be normalized to $\Omega_0$ according to:
\begin{equation}
\Omega_{n_i,n_{1,i}}^{n_f,n_{1,f}}(n_\pi)=\Omega_0
\frac{\Upsilon_{n_i,n_{1,i}}^{n_f,n_{1
,f}}}{\Upsilon_{50,0}^{51,1}}=\Omega_0
\frac{\mathcal{M}_{n_i,n_{1,i}}^{n_f,n_{1
,f}}}{\mathcal{M}_{50,0}^{51,1}}.
\end{equation}
We were therefore able to construct the matrix of a Hamiltonian at
constant $m$ including $M$ multiplicities in $n$ (ie.
$n=m+1,\cdots,m+M$), and $2N+1$ multiplicities in $n_\pi$ (ie.
$n_\pi-N,\cdots , n_\pi ,\cdots , n_\pi+N$). For the initial level
positions $\mathcal{E}(k;E)$ we use the quadratic Stark expansion of
Eq.~\ref{eq:stark0-2} for all levels apart from $\ket{g}$,
$\ket{e}$, $\ket{i}$ and $\ket{i^\prime}$, as explained above. After
diagonalization of this Hamiltonian, we are left with a new ladder
of levels, $\widetilde{\mathcal{E}}(k;E)$. These levels are slightly
shifted in energy with respect to the unperturbed states and it is
of course this shift that we hope to use in order to render
$\widetilde{\mathcal{E}}(g,n_\pi;E)$ and
$\widetilde{\mathcal{E}}(e,n_\pi;E)$ parallel. This shift being
small, and there being no resonant coupling due to the dressing
field, we are able to associate each of the dressed levels
unambiguously with one of the unperturbed levels. We label the state
arising from $\ket{n,n_1,m,n_\pi}$ as
$\widetilde{\ket{n,n_1,m,n_\pi}}$ where the new quantum numbers
characterize the now mixed atom-field state. We computed the change
in energy of $\widetilde{\mathcal{E}}(g,n_\pi;E)$ and
$\widetilde{\mathcal{E}}(e,n_\pi;E)$ as one increases $M$ or $N$ by
one, hence increasing the size of the Hamiltonian diagonalized. We
found this difference to decrease exponentially and fall below the
Hz level for $M \geq 6$ and $N \geq 4$. By a diagonalization of the
Hamiltonian with $M=6$ and $N=4$ we were therefore able to calculate
$\widetilde{\mathcal{E}}(g,n_\pi;E)$ and
$\widetilde{\mathcal{E}}(e,n_\pi;E)$ to the Hz level for a given
electric field $E$, microwave power (corresponding Rabi frequency
$\Omega_0$) and detuning $\delta_0$. We can therefore write:
\begin{equation}
\mathcal{E}_k(E,\Omega_0,\delta_0)=
\widetilde{\mathcal{E}}(k,n_\pi;E)-n_\pi\hbar\omega_0,
\label{eq:Epothabille}
\end{equation}
for the potential energy of a trapped atom in state $k=g$ or $e$, in
the presence of the microwave dressing field.

From these energies one can determine the dressed level transition
frequency:
\begin{equation}
\omega_{\tilde{e}\tilde{g}}(E,\Omega_0,\delta_0)=
\left[\mathcal{E}_{e}(E,\Omega_0,\delta_0)-
\mathcal{E}_{g}(E,\Omega_0,\delta_0)\right]/\hbar,
\end{equation}
and analyse its dependence on the electric field. In the
neighbourhood of $E_a$ we have:
\begin{eqnarray}
\omega_{\tilde{e}\tilde{g}}(E,\Omega_0,\delta_0)&=&
\omega_{\tilde{e}\tilde{g}}(E_a,\Omega_0,\delta_0)\\
&&+ L(E_a,\Omega_0,\delta_0)(E-E_a)\nonumber\\
&&+ Q(E_a,\Omega_0,\delta_0)(E-E_a)^2\nonumber\\
&&+ O((E-E_a)^3),\nonumber
\end{eqnarray}
and it is the coefficients $L(E_a,\Omega_0,\delta_0)$ and
$Q(E_a,\Omega_0,\delta_0)$ that we hope to cancel, as we did under
the simpler, two-level treatment laid out in
Sec.~\ref{ssec:leveldressing}. Unfortunately, due to the effect of
the microwave field upon level $\ket{e}$ as well as $\ket{g}$, we
find that, while it is possible for any $\delta_0$ to find an
$\Omega_0$ such that $L(E_a,\Omega_0,\delta_0)=0$, it remains
nonetheless impossible to simultaneously cancel
$Q(E_a,\Omega_0,\delta_0)$. We therefore content ourselves with
canceling $L(E_a,\Omega_0,\delta_0)$ and bringing
$Q(E_a,\Omega_0,\delta_0)$ close to its minimum for the reasonable
parameter values $\Omega_0=2\pi \times 200.000$~MHz and
$\delta_0=2\pi \times 555.907$~MHz. The remaining frequency
dispersion corresponding to the field excursion $\Delta E=1$~V/m is
of the order of 10~Hz.

Turning now to the second item on our list of imperfections in the
simplistic treatment of Sec.~\ref{ssec:leveldressing}, we must
analyse the effect of the variation of $\Omega (\vec{r},t)$ due to
the mode profile, the angle $\theta (\vec{r},t)$ or, indeed, simple
technical noise on the microwave amplifier. By analysing the
variation of $\omega_{\tilde{e}\tilde{g}}(E_a,\Omega_0 + \Delta
\Omega,\delta_0)$ with $\Delta \Omega$ we were able to determine
that the consequent inhomogeneous broadening is less than $10$~Hz
for $\Delta \Omega / \Omega <2\times10^{-7}$. This sets a tight
condition on the power stability of the microwave source for the
dressing. Both the excursion of the atom in the microwave field
profile and the variation of $\theta (\vec{r},t)$ introduce a
$\Delta \Omega$ smaller than this.

The rate of spontaneous cascades of the type
$\widetilde{\ket{g,n_\pi}} \rightarrow \widetilde{\ket{k,n_\pi-1}}$
is found by calculating the dipole
$\vec{D}(\widetilde{g}\widetilde{k})$ between
$\widetilde{\ket{g,n_\pi}}$ and all the eigenstates
$\widetilde{\ket{k,n_\pi^\prime}}$ of lower energy and then summing
the rates of spontaneous transitions, proportional to the square of
the dipole. The lifetimes thus calculated converge rapidly with
increasing $M$ and $N$ towards 11.9~s for
$\ket{\widetilde{g},\widetilde{n_\pi}}$ and 62.0~s for
$\ket{\widetilde{e},\widetilde{n_\pi}}$ (much longer because of the
greater detuning of the microwave for transitions within the $m=50$
multiplicity and the consequently smaller contamination of $\ket{e}$
by other levels). These lifetimes are much longer than those already
imposed by other limitations as presented in
Sec.~\ref{ssec:spontinhibimperf} and are therefore not an obstacle.

Finally, it was checked that there were no resonances between the
dressed circular states and adjacent levels of different $m$ due to
the small $\sigma$-polarized component of the microwave field. No
transition was found towards the multiplicities of $\Delta m = \pm\
1$ with a detuning of less than 100~MHz. These transitions
consequently pose no problem. All transitions at $\Delta m = \pm\
2,3\ldots$ have Rabi frequencies  $\ll 1$~Hz, these processes being
of second or higher order, passing necessarily by a virtual
transition at $\Delta m = \pm\ 1$ which we have just seen to be
hugely nonresonant, and hence strongly suppressed.

\subsection{Simulation of the coherence control}\label{ssec:dressingperformance}

In order to estimate the coherence decay time $T_2$ we simulated a
Ramsey interferometry experiment in the presence of the dressing
microwave. We simulate the trajectories of an atom as in
Sec.~\ref{ssec:trapsimulations} with the state dependent potential
energy given by:
\begin{equation}
\mathcal{E}_k(\vec{r},t)=\overline{\mathcal{E}}_k(E(\vec{r},t),\Omega_0
f(\vec{r}) \cos \theta(\vec{r},t), \delta_0),
\end{equation}
where the $\overline{\mathcal{E}}_k(E,\Omega,\delta)$ is a series
expansion of the $\mathcal{E}_k$ to fourth order in $E$ and $\Omega$
around $(E_a,\Omega_0,\delta_0)$. $E(\vec{r},t)$ and
$\theta(\vec{r},t)$ are the electric field amplitude and tilt, and
$f(\vec{r})$ accounts for the dressing microwave spatial mode (see
Sec.~\ref{ssec:leveldressing}).

\begin{figure}
\begin{center}
\resizebox{0.75\columnwidth}{!}{
\includegraphics{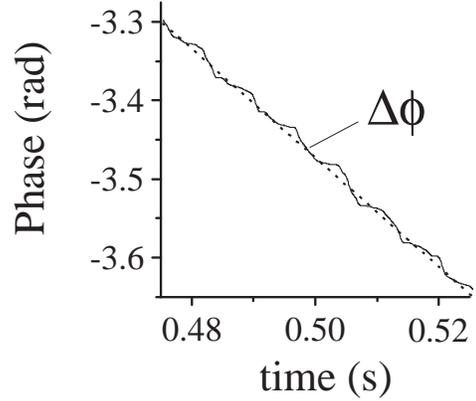}
}
\end{center}
\caption{\label{fig:phaseacquired} Phase acquired by an
$\ket{\widetilde{e}}/\ket{\widetilde{g}}$ coherence over 40~ms. The
trajectory considered is in trap A, under the conditions of
Table~\ref{tab:performance} and at $T_0=300$~nK.}
\end{figure}

We assume that the atom undergoes two short microwave $\pi/2$
pulses, resonant on the $\ket{\widetilde{g}}/\ket{\widetilde{e}}$
transition (frequency
$\omega_{\tilde{e}\tilde{g}}(E_a,\Omega_0,\delta_0)$) at times $t_i$
and $t_f$, separated by a long delay $T=t_f-t_i$. For each state we
follow the state dependent trajectory. For atoms taken from a sample
at $0.3\ \mu$K, the average separation between the two is
$\sim20$~nm, much smaller than the de Broglie wavelength
$\lambda_{DB}=0.34\ \mu$m. The Stern Gerlach effect can therefore be
neglected. The dephasing between the two states is then due only to
the phase $\phi(T)$ acquired by the
$\ket{\widetilde{e}}/\ket{\widetilde{g}}$ coherence along the
trajectory $\vec{r}(t)$:
\begin{eqnarray}
\phi(T) & =  \int_{t_i}^{t_f}   \left[ \right.&
\overline{\mathcal{E}}_{e}(E\left[\vec{r}(t),t\right],\Omega_0
f\left[\vec{r}(t)\right] \cos
\theta\left[\vec{r}(t),t\right], \delta_0)  \nonumber \\
& & - \overline{\mathcal{E}}_{g}(E\left[\vec{r}(t),t\right],\Omega_0
f\left[\vec{r}(t)\right] \cos
\theta\left[\vec{r}(t),t\right], \delta_0) \nonumber \\
& & \left. - \omega_{\tilde{e}\tilde{g}}(E_a,\Omega_0,\delta_0)\
\right] dt.
\end{eqnarray}

\begin{figure}
\begin{center}
\resizebox{0.85\columnwidth}{!}{
\includegraphics{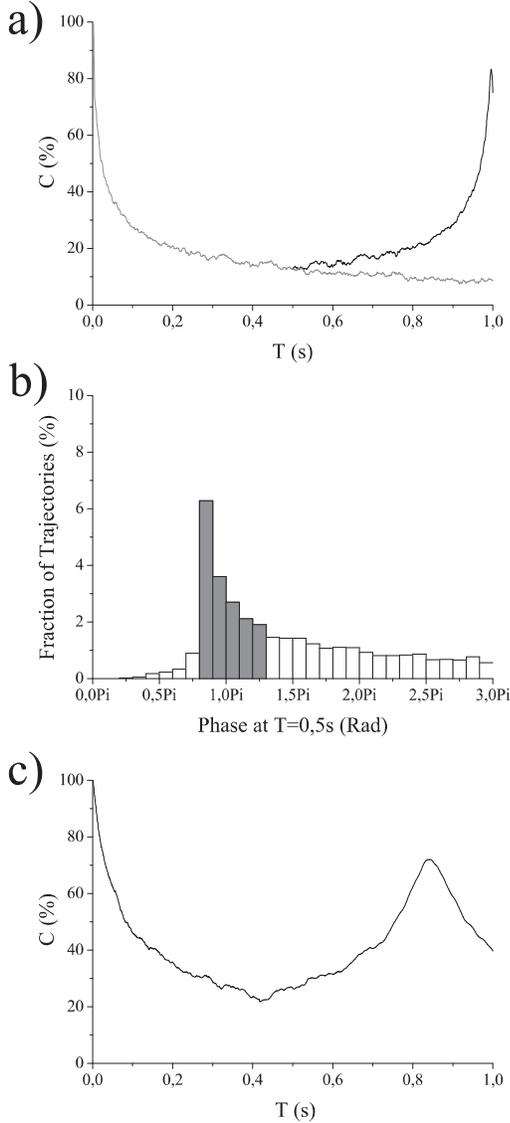}
}
\end{center}
\caption{\label{fig:CvsT} (a) Simulated Ramsey fringe contrast $C$
as a function of the time interval $T$ between two $\pi/2$ pulses in
the case of Trap A. Conditions are as in Table~\ref{tab:performance}
with $T_0=300$~nK, we average over 10000 trajectories. Grey line:
contrast decay without echo. The contrast undergoes a
non-exponential decay, falling to 50\% in 24 ms, and to 13\% in 0.5
s. Black line: contrast decay and then revival after a $\pi$-pulse
at $T_\pi=0.5$~s, reaching 83\% at $T=1$~s. (b) The histogram of the
phases accumulated at $T$=0.5 s sheds light on this behaviour. A
subset of the trajectories (white bars in the histogram) accumulates
large phases and its contribution averages rapidly to zero. The
other trajectories (grey bars), remaining closer to the centre of
the trap, have a much slower phase drift and account for the slow
decay at long times. (c) Simulated Ramsey fringe contrast $C(T)$ in
Trap B under the conditions of Table~\ref{tab:performance} and with
$T_0=300$~nK, averaged over 2500 trajectories. Here we have applied
a $\pi$-pulse at $T=420$~ms and we see a revival of the contrast
reaching 72\% at $T=840$~ms.}
\end{figure}

Fig.~\ref{fig:phaseacquired} represents this phase evolution for a
given trajectory over a 40~ms time interval. One can see that, over
times longer than the trap oscillation period, the evolution is
quite linear. On times scales of the order of the trap period,
$\phi(T)$ proceeds by small steps of amplitude $\Delta\phi$ which
occur when the atom is furthest from the origin. The final fringe
contrast for the whole set of trajectories can be evaluated by
$C=(\overline{\cos \phi}^2 +\overline{\sin \phi}^2)^{1/2}$ (where
the bar denotes an average over all trajectories). The time
evolution of $C$, shown in Fig.~\ref{fig:CvsT}(a), provides an
estimate of $T_2$ at around 25~ms. This is in qualitative agreement
with the predictions of Sec.~\ref{ssec:dressingimperf}. The
evolution of the contrast over long time is, however,
nonexponential. After a very rapid initial fall, $C(T)$ decays at a
much slower rate. We attribute this behaviour to the fraction of
trajectories which remain very close to the center of the trap and
whose phase drift can be much smaller than 1 over significantly long
times [see Fig.~\ref{fig:CvsT}(b)].

As the phase acquired is almost perfectly linear with time for all
trapped trajectories, one can try to combat the long term phase
spreading using an echo technique reminiscent of photon-echo
techniques. Coherence preserving echoes have also been tested for
trapped ground-state atoms or
ions~\cite{TR_ECHODAVIDSON03,TR_ECHOBONN03,ION_WINELANDTRANSPORTION02}.
In order to examine the efficiency of this technique we simulate, at
a time $T_{\pi}$ after the first Ramsey pulse, an additional
$\pi$-pulse on the $\ket{\widetilde{e}}/\ket{\widetilde{g}}$
transition, exchanging the populations of the two states. Pulse
imperfections are taken into account for by considering, instead of
a perfect $\pi$-pulse, a pulse performing the following
transformation:
\begin{eqnarray}
\ket{\widetilde{g}} & \rightarrow & \cos (\frac{\pi + \vartheta}{2})
\ket{\widetilde{g}} + \sin (\frac{\pi + \vartheta}{2}) \ket{\widetilde{e}},\\
\ket{\widetilde{e}} & \rightarrow & -\sin (\frac{\pi +
\vartheta}{2}) \ket{\widetilde{g}} + \cos (\frac{\pi +
\vartheta}{2}) \ket{\widetilde{e}}.
\end{eqnarray}
where $\vartheta$ can be considered as the error in the angle of
rotation in the Bloch sphere, $\vartheta = 0$ corresponding to the
ideal case. The $\vartheta$ for each trajectory is randomly chosen
from a gaussian distribution, centered around $0$, with a dispersion
of 0.1 Rad. In the ideal case where $\vartheta=0$ the phase of the
$\ket{\widetilde{e}}/\ket{\widetilde{g}}$ coherence is multiplied by
-1 upon application of the $\pi$-pulse. During the subsequent
evolution, the phase drift continues as before and thus returns
towards zero. At time $T=2T_{\pi}$, all phases are zero to within an
uncertainty of the order of the average phase step amplitude
$\Delta\phi$.

Fig.~\ref{fig:CvsT}(a) also presents the contrast $C$ obtained under
these conditions as a function of $T$ with a $\pi$-pulse applied at
$T_\pi=0.5$~s. We see it increase sharply up to 83\% around
$T=2T_{\pi}=1$ s. This very high contrast corresponds to an
effective $T_2=5.4$ s. Even for atoms at $T_0=1\ \mu$K, we obtain
$C(2T_{\pi})=$57.3\%. Fig.~\ref{fig:CvsT}(c) shows that the
technique also works in trap B, producing an effective $T_2$ of
2.6~s. More complex echo sequences can be envisaged to improve the
final Ramsey fringe contrast. Ideal $\pi$-pulses repeated at shorter
time intervals can maintain the coherence over time scales in the
minute range, being only limited by the radiative lifetime.

\section{Conclusions and perspectives}\label{sec:conclusion}

We have shown that it is possible to integrate on a microchip the
elements necessary to excite a single atom into a circular Rydberg
state, to trap it, and to perform coherent manipulation of its
internal states over seconds. The wealth of possible electrode
geometries makes it easy to extend these results to the case of
atomic waveguides or arrays of traps. We have also obtained
preliminary results proving that it could be possible to integrate
the detection of the Rydberg states on the chip itself. This relies
on the usual state dependent field ionization of the atom followed
by detection of the electron by a thin superconducting wire, such as
those already used in fast photon detectors~\cite{SEMENOVREVIEW_02}.
Our proposal would offer a complete ``toolbox'' to study and control
a single quantum system in an extremely versatile manner. It also
forms a scalable architecture for quantum information processing.
The experimental realization of this project will require the
adaptation of atom-chip techniques from room temperature to the
superconducting regime. We are currently developing a cryogenic
experiment able to achieve this, opening new perspectives for
atom-chip experiments.

\begin{acknowledgement}
Laboratoire Kastler Brossel is a laboratory of Universit\'{e}
Pierre et Marie Curie and ENS, associated to CNRS (UMR 8552). We
acknowledge support of the European Community (QUEST and QGates
projects), of the Japan Science and Technology corporation
(ICORP Program: Quan\-tum
Entanglement), and of the French Ministry of research (programme ACI).
\end{acknowledgement}
% BibTeX users please use
%\bibliographystyle{toto}
%\bibliography{txt,qc,qi,mx,ens,refmanquantes}

\end{document}